\documentclass[12pt]{iopart}

\usepackage{graphicx}

\usepackage{cite}

\begin{document}


\author{D.~E.~Kahana$^1$ and S.~H.~Kahana$^2$}
\address{1. Physics Department, Brookhaven National Laboratory\\
   Upton, NY 11973, USA}
\address{2. 31 Pembrook Drive, Stony Brook, NY 11790}
\eads{\mailto{kahana@bnl.gov}}
\title{Higgs and Top Masses from Dynamical Symmetry Breaking - Revisited}

\date{\today}  

\begin{abstract} 

We re-examine our former predictions ~\cite{kahanath1,kahanath2} of the top
and Higgs masses via dynamical symmetry breaking in a 4-fermion theory which
produces the Higgs as a bound state, and relates the top and Higgs masses to
$m_W$. The use of dynamical symmetry breaking was stongly motivated by the
apparent equality, within a factor of two, of the known and expected masses of
the $W$, $Z$, top and Higgs. In later work ~\cite{kahanath2} we evaluated the
masses self-consistently at the mass-poles, which resulted in predictions of
$m_t \sim 175$ GeV, and $m_H \sim 125$ GeV as central values within ranges
produced by varying the measured strong coupling. Figures~(1) and (2) result
from evolution down to $m_W$ while the number quoted for the top quark mass,
{\it i.e.} 175 GeV includes an evolution back up to the top and use of the
determination of $\alpha_s$ at LEP at that time. $m_H$ is less dependent on
the value of the strong coupling. The variation of the predicted masses for a
range of the strong and electro-weak couplings $\alpha_s$, $\alpha_W$ at $m_W$
are exhibited in Figure~(3) and Figure~(4) reproduced from the last
work~\cite{kahanath2}, which was submitted to PRD well before the first FNAL
publications~\cite{CDF1,D0top} suggesting evidence for the top.

\end{abstract} 
\maketitle
 
\def\ff{$4$-fermion }

\def\ff{$4$-fermion }
\def\ref#1{[{\bf #1}]}
\section{Introduction}
In this comment we recall predictions for the top and Higgs masses made in an
earlier work on dynamical symmetry breaking~\cite{kahanath2}. Earlier
authors~\cite{bjorken,bardeen,terazawa,eguchi} have exploited this subject.
The specific \ff model of dynamical symmetry breaking
presented~\cite{kahanath1,kahanath2} is an NJL-like~\cite{NJL} 4-fermion
theory but crucially extended to include electro-weak current-current vector
interactions~\cite{dkahana1}. The theory is imagined to be valid at some high
scale $\mu$, presumably an effective theory arising from new physics above
$\mu$. Variation of this scale by four orders of magnitude from $10^{10}$ to
$10^{14}$ GeV has no appreciable effect on the mass predictions (see
Figure~(2)). This scale then acts as an effective cutoff heralding the
entrance for new physics, probably well above $\mu$. Again, no explanation is
offered for the number and character of the model's fundamental fermions, nor
for the large disparity in mass scales, i.e. $m_f(\mu)\ll \mu $.  Rather, a
central point of our previous calculation was that composite bosons, the W, Z
and Higgs, with masses near $2 m_f(\mu)$ arise naturally in the theory.  

These are just fermion--antifermion bound states produced by the \ff
interaction. This phenomenon is well described in the papers of Nambu and
Jona--Lasinio~\cite{NJL} who were themselves evidently inspired by the
superconducting theory of Bardeen, Cooper and Schreiffer~\cite{BCS}, The vital
element of the model's inherent dynamic (chiral) symmetry breaking is the
specific relation that emerges between the fundamental quark masses,
especially the top, and the induced, composite, boson masses.

           
Previously~\cite{kahanath1,kahanath2} we abstracted simple, asymptotic mass
relationships from the \ff theory, and used these as boundary conditions on
the standard model renormalisation group evolution (RGE) equations. This was
done at a matching scale $\mu$ somewhat below the GUTS scale, where the
electroweak (EW) sector can still be treated as approximately independent of
QCD ($SU(3)_c$). Values for the top and Higgs masses then followed from
downward evolution of the top-Higgs and Higgs-self couplings to scales near
and above $m_W$, assuming no new physics intervened between the upper and
lower scales.  The W, Z and Higgs, with masses near $2 m_t(\mu)$ arise
naturally in the theory.

Since the scale $\mu$ at which any new physics enters is so high, the theory
becomes a weak coupling, albeit specially constrained, version of the Standard
Model for lower scales right down to so-far attainable experimental
energies. The stong interactions, absent in the initial action, enter, as
indicated above, through the RGE equations.  Finally we discuss interesting
possible nodifications to the model, the simplest of which does not
disturb the stability of the predictions, and a second which may enable the
introduction of new physics. We begin by briefly summarising the previous
development.

\section{Recap}
The model is defined by the Lagrangian:

\begin{eqnarray}
{\cal L} & =\bar\psi i(\gamma \cdot \partial) \psi - {1\over 2}
  [(\bar\psi G_S \psi)^2-(\bar\psi G_S \tau\gamma_5\psi)^2]\nonumber \\
 & - {1\over 2} G_B^2(\bar\psi\gamma_\mu Y \psi)^2-{1\over 2}
      G_W^2(\bar\psi\gamma_\mu \tau P_L\psi)^2
\end{eqnarray}

The field operator is $\psi=\lbrace
f_i \rbrace$, and the index $i$ runs over all fermions,
$i=\lbrace(t,b,\tau,\nu_{\tau}),(c,s,...),...\rbrace$. The scalar-coupling
matrix $G_s$ is taken diagonal and the dimensionful couplings are adjusted to
produce the known fermion masses dynamically; in practice only the top
acquires an appreciable mass. The model admits bound states corresponding
to the Higgs as well as the gauge bosons of the standard electroweak theory,
and is essentially equivalent to the Standard Model below some high mass
scale $\mu$. It is the scalar  which permit a gap equation and which
generate the composite Higgs, while the vector currents play
a similar role for the $W,Z$, all with masses of the order of $m_t$, 

The construction of the   effective actions for both scalar and vector
sectors is laid out in the original work~\cite{kahanath1}. Following
D.~E.~Kahana~\cite{dkahana1} as well as Gross and
Neveu~\cite{grossneveu}, a classical Lagrangian is introduced containing
auxilliary fields $\sigma,\pi,B_{\mu},W_{\mu}$. The latter fields are
appropriately shifted so as to elimate the four-fermion terms in Eq.1,
leaving scalar and vector couplings of the fermion and bosons but devoid of
boson kinetic and mass components. A derivative expansion of the
resulting effective action generates  the usual NJL(BCS) scalar gap
equation, at first order, and composite boson masses, scalar Higgs and
vectors $W$ and $Z$, at second order. Higher orders 
complete the usual Standard Model abelian and non-abelian  actions.

A necessary fine tuning of the scalar coupling, for an assumed diagonal
coupling matrix $G_s$ yields at second order in fields, the Higgs mass
formula:
\begin{equation}
m^2_H(\mu) = 4\sum_f m^2_f(\mu)
\end{equation}
\noindent  implying the Higgs is a composite of all $f\bar f$ pairs.

Bound states also exist in the vector sector corresponding to the $W$, $Z$,
and the photon. A similar fine tuning of the vector coupling is required, but
here with the added physical requirement that the photon mass vanish, leading
at lowest order in the electroweak and Yukawa couplings to the mass
relationship
\begin{equation}
m^2_W(\mu) = \left(\frac{3}{8} \sum_f m^2_f(\mu)^2 +\frac{1}{8}\sum_l
m^2_l(\mu)
\right)
\end{equation}

This equation contains a factor $n_c/n_g$ which evaluates to {\it e.g.} 3/4
for three colors and four generations, distinguishing it from the 3/8 that
appears in the following relation for the weak angle.

Diagonalisation of the neutral vector boson action in each generation
by rotating ($W_0,B$) into ($Z,A$) results in
\begin{equation}
\sin^2 (\theta_W(\mu)) = (\sum_i(Q_i)^2)^{-1}=\frac{3}{8}
\end{equation}
with the denominator on the right hand side of the latter equation being
summed over the charges $Q_i$ in one generation, lending a physical meaning to
the oft cited SU(5) Clebsch coefficient, that determines the weak angle in the
minimal SU(5) GUT.

The dimensionful couplings of the \ff theory are replaced, after fine-tuning
and wave function renormalisation, by the dimensionless couplings of the
Standard Model~\cite{marciano1,marciano2}, and the gradient expansion of the
effective action is in fact an expansion in these dimensionless electroweak
couplings. One has for the scalars
\medskip
\begin{eqnarray}
g_S &= G_S Z_S^{-\frac{1}{2}},\nonumber \\ 
Z_S &= \frac{1}{2} Tr \left[G_S^2\frac{1}{(\partial^2 + M^2)^2}\right],
\end{eqnarray}

\noindent Similarly, for the vector couplings one has
\begin{equation}
\frac{g_2}{2} = \frac{G_W}{\sqrt{Z_W}} \&
\frac{g'}{2} = \frac{G_B}{\sqrt{Z_B}}
\end{equation}
\noindent where the usual relationship obtains between $g_2$ and $g'$
\begin{equation}
g_2 \sin (\theta_W) = g' \cos (\theta_W).
\end{equation}

\subsection{Renormalisation Group Evolution}

From equations $\lbrace (2), (4), (5)\rbrace$, valid presumably at a scale
$\mu$ where the cross coupling between the EW and strong sectors is small but
still well below the cutoff $\Lambda$, we derived values for the top and Higgs
masses at a scale near $m_W$. The theory leading to these equations is
equivalent to the electroweak sector of the Standard Model below $\mu$, and
the framework for connecting the scales $\mu$ and $m_W$ is provided by the
Standard Model RGE. So, $SU(3)_c$ influences on the top and Higgs masses are
included through the renormalisation group, below the matching scale
$\mu$. Defining:
\begin{equation}
\kappa_t = \frac{g_t^2}{2\pi},
\end{equation}
\noindent one has
\def\kt{\kappa_t}
\def\as{\alpha_S}
\def\aw{\alpha_W}
\def\aone{\alpha_1}
\begin{equation}
\frac{d\kt}{dt} = \frac{9}{4\pi} \kt^2 - \frac{4}{\pi} \kt \as
- \frac{9}{8\pi} \kt \aw - \frac{17}{4\pi} \kt \aone,
\end{equation}
with $\as$, $\aw$, $\aone$ taken equal to $\alpha_3$, $\alpha_2$, $\alpha_1$,
respectively, as in reference~\cite{marciano1,marciano2}, and where
$t=\ln(\frac{q}{m})$. 

\noindent With these choices one finds
\begin{equation}
m_t = \frac{1}{\sqrt{2}} g_t v,\\
m_W = \frac{g_W}{2} v,\\
\end{equation}
\noindent  where $v$ is the standard EW vev.

\noindent Also, taking $m_H^2 = 2\lambda v^2$ the evolution equation for the Higgs
self-coupling is, to the same (one-loop) order~\cite{dawson}:

\begin{equation}
\frac{d\lambda}{dt} = \frac{1}{16\pi^2} \left\lbrace 12 \lambda^2 + 6 \lambda g_t^2
- 3g_t^4 - \frac{3}{2} \lambda  \left(3 g_W^2 + {g'}^2\right)
+ \frac{3}{16}
\left(2 g_W^4 + \left(g_W^2 + {g'}^2\right)^2\right)\right\rbrace.
\end{equation}

\noindent Redefining the standard choice of couplings \cite{marciano1}
\begin{equation}
\aone =\frac{5}{3} \alpha', \alpha_1 = \frac{g_1^2}{4\pi}, \alpha' = \frac{{g'}^2}{4\pi}.
\end{equation}
\noindent and setting
\begin{equation}
\sigma = \frac{\lambda}{4\pi}.
\end{equation}
\noindent results in
\begin{equation}
\frac{d\sigma}{dt} = \frac{1}{2\pi} \left\lbrace 
12 \sigma^2 + 6 \sigma \kt
- 3\kt^2 
- \frac{9}{2} \sigma 
\left(\alpha_W + \frac{1}{5} \alpha_1\right)
+ \frac{3}{16} 
\left(2 \alpha_W^2 +
\left(\alpha_W + \frac{3}{5} \alpha_1 \right)^2
\right)
\right\rbrace.
\end{equation}

The latter second order mass relations impose boundary conditions on the
differential equations for $\lambda$ and $\sigma$ at the scale $\mu$. These
are, to lowest order:
\begin{equation}
m_t^2 = \frac{8}{3}\,m_W^2 
\end{equation}
and
\begin{equation}
m_H^2 = 4 m_t^2 = \frac{32}{3} \,m_W^2.
\end{equation}

\subsection{Solution of the RG Equations.}

It is possible to obtain an explicit solution to the differential
equation for $\frac{d\kappa_t}{dt}$, and a perturbative
solution for  $\frac{d\sigma_t}{dt}$~\cite{kahanath2}. The exact solution 
for  $\kappa_t$ involves an integration constant $D$ given by

\begin{equation}
D = \frac{1}{\kt(0)},
\end{equation}
\noindent and directly yields the running top mass at the scale $m_W$ from
\begin{equation}
m_t^2(m_W) = \frac{2\kt(0)}{\alpha_W (0)} m_W^2(m_W).
\end{equation}
\noindent To self-consistently determine the physical top mass as a pole in the
top quark propagator, one must then run $m_t(m_W)$ back up to get $m_t(m_t)$.

The cross coupling in Eq(10) complicates its 
solution. The pure scalar self-coupling result 
\begin{equation}
\sigma_0 (t) = \frac{\sigma_0(0)}{1 - \frac{6}{\pi} \sigma_0(0) t},
\end{equation}
\noindent may be improved  perturbatively
\begin{equation}
\sigma (t) = \sigma_0 (t) + \sigma_1 (t).
\end{equation}
The contribution from $\sigma_1$ is however small as exhibited in
Figure~(1)~\cite{kahanath2}.

Results from numerical integration of the equations down to $m_W$ are
displayed in Table 1, and Figures(1--4). We have varied the inputs to these
calculations, the strong and electroweak couplings $\alpha_{i0}$, $i =
1,W,S$ over a reasonable range(see comments in the Abstract), The W mass is
fixed at 80.1 GeV somewhat lower than the presently accepted value.  There are no free
parameters in the theory, the couplings and $m_W$ being determined from
experiment. A possible exception is the upper cutoff $\Lambda$, which is surely
well above $\mu$ and has essentially no effect on $m_t$ and $m_H$. Any
dependence other than logarithmic on $\Lambda$ has been eliminated by fine
tuning, while residual $\ln (\Lambda)$ presence is transmuted into
dependence on the dimensionless couplings.

The effect of imposing boundary conditions sharply at a scale $\mu$ remains
to be examined. As we noted above, $\mu$ is that point, when one is evolving
downward in mass, at which the $g_i$ become interdependent.  For example, the
top quark evolution is strongly influenced by $SU(3)_c$ from $\mu\sim 10^{14}
$ downward, and the running of $\alpha_W$ is also significant.  Varying $\mu$
over four orders of magnitude from $\mu = 10^{10}$ GeV to $\mu = 10^{14}$ GeV
has practically no effect on $m_t$, and only a small effect on $m_H$. This
remarkable result is demonstrated in Fig~(1) for a range of the
couplings, and lends credence to our use of a sharp boundary condition.

The one physical parameter sensitive to $\mu$ is the weak mixing angle
$\theta_W$. We indicated \cite{kahanath1} that, for one loop evolution,
$\sin^2(\theta_W)$ achieves its experimental value $\sim 0.23$ at $m_W(\mu)$
for $\mu\sim 10^{13}$ GeV. Unlike GUTS, the present theory need not have a
single scale at which the gauge couplings are equal. There is a unification
present in this model simply implying that the Standard Model should evolve
smoothly into the effective \ff theory when the couplings become weak.  Table
(1) displays the value of the couplings at scale $\mu$; the $\alpha_i$ are the
experimental values determined at $m_W$ evolved upward to $\mu$ at 1-loop and
$\kt(\mu)$ is obtained from the boundary condition $\frac{\kt}{\alpha_2} =
\frac{4}{3}$.  It is clear that the couplings are indeed all small at $\mu$,
again justifying the placing of the boundary conditions there.

\section{Conclusions}
\medskip

Figures~(3)~and~(4) show the variations of $m_t$ and $m_H$ with the strong and
electroweak couplings, respectively. The strong coupling is less well
known. Using as central values $\alpha_{S0} = 0.107$, $\alpha_{W0}=0.0344$, we
get the aforementioned central values $m_t\simeq 175$ GeV and $m_H\simeq 125$
GeV which remain valid after evolution to the pole masses.  Further small
contributions to Eq(19), from non-leading log terms in defining the top pole
and from running the W mass, more or less cancel.

In summary, one gets remarkably stable predictions for the top and Higgs
masses and in a parameter free fashion. The only inputs were the
experimentally known couplings and the W-mass. A characteristic prediction of
this type of theory is $m_h<m_t$, so that the Higgs, which is in our model
a condensate of all elemental fermions, is deeply bound.

It would appear this note and its contained recollections are
particularily timely in view of present activity at the CERN LHC. We are
in fact in the midst of extending the model to include a second
Higgs doublet and will soon report on this. We indicated that such an
enterprise necessarily includes other fermions, notably the bottom. The
presence of the latter   has little direct effect on the top mass,
but such a toy model may very well provide information on the 2-Higgs sector.

We should not end the discussion without mention of supersymmetry. Our
direction forward, introducing two Higgs doublets, has been chosen because of
the effects of SUSY. The latter in its unbroken form preseves the very chiral
symmetry, the breaking of which, after all, yields all of our results.
Moreover the early efforts of Buchmuller and Love~\cite{buchmuller} in
demonstrating the apparent incompatibility of SUSY and dynamical chiral
symmetry breaking, deserve recognition. One could, of course, break SUSY in
the presently accepted "soft" form but not without loss of the naturalness of
dynamical chiral symmetry breaking. We must await the final word on the
existence or non-existence of supersymmetry from the LHC.

\section{Acknowledgements}

This  manuscript  has  been  authored  under  the  US  DOE  grant
NO. DE-AC02-98CH10886.
The authors appreciate greatly the many early conversations with Bill
Marciano, which provided both needed education and encouragement, as well as
those with Frank Paige on high scales.
One of the authors (SHK) would also like to thank the Alexander von Humboldt
Foundation (Bonn, Germany) for partial support throughout the long history of
the work.

\section*{References}
\bibliography{thf1}
\bibliographystyle{iopart-num}

\begin{figure}
\includegraphics*[scale=0.8]{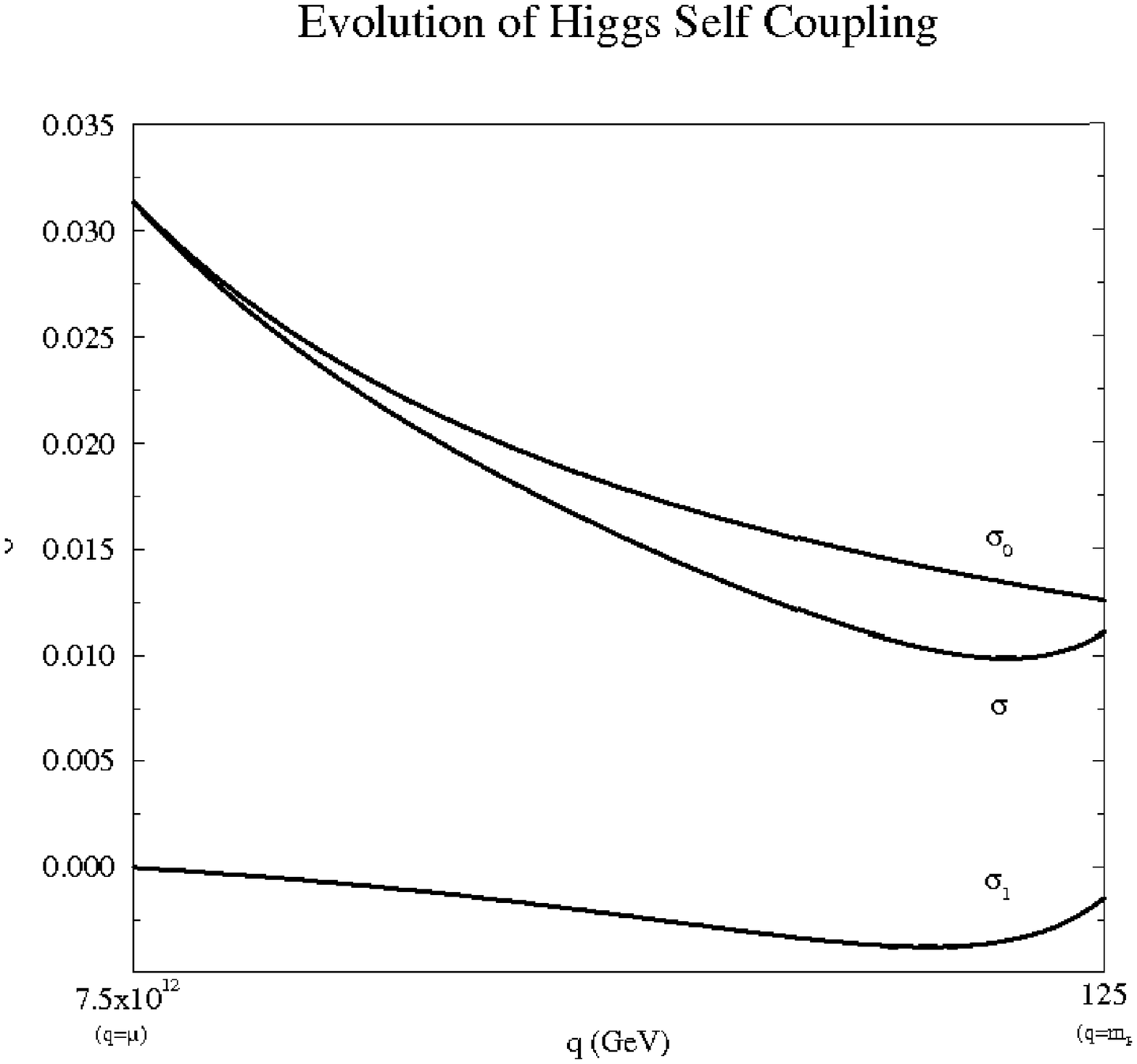}
\caption[]{Evolution of the reduced Higgs Self Coupling $\sigma =
  \sigma_0+\sigma_1$ over the range from $m_W$ to $\mu=10^{14}$. The
  perturbation $\sigma_1$ remains small.}
\label{fig:Fig.(1)}

\end{figure}
\clearpage

\begin{figure}
\includegraphics*[scale=0.8]{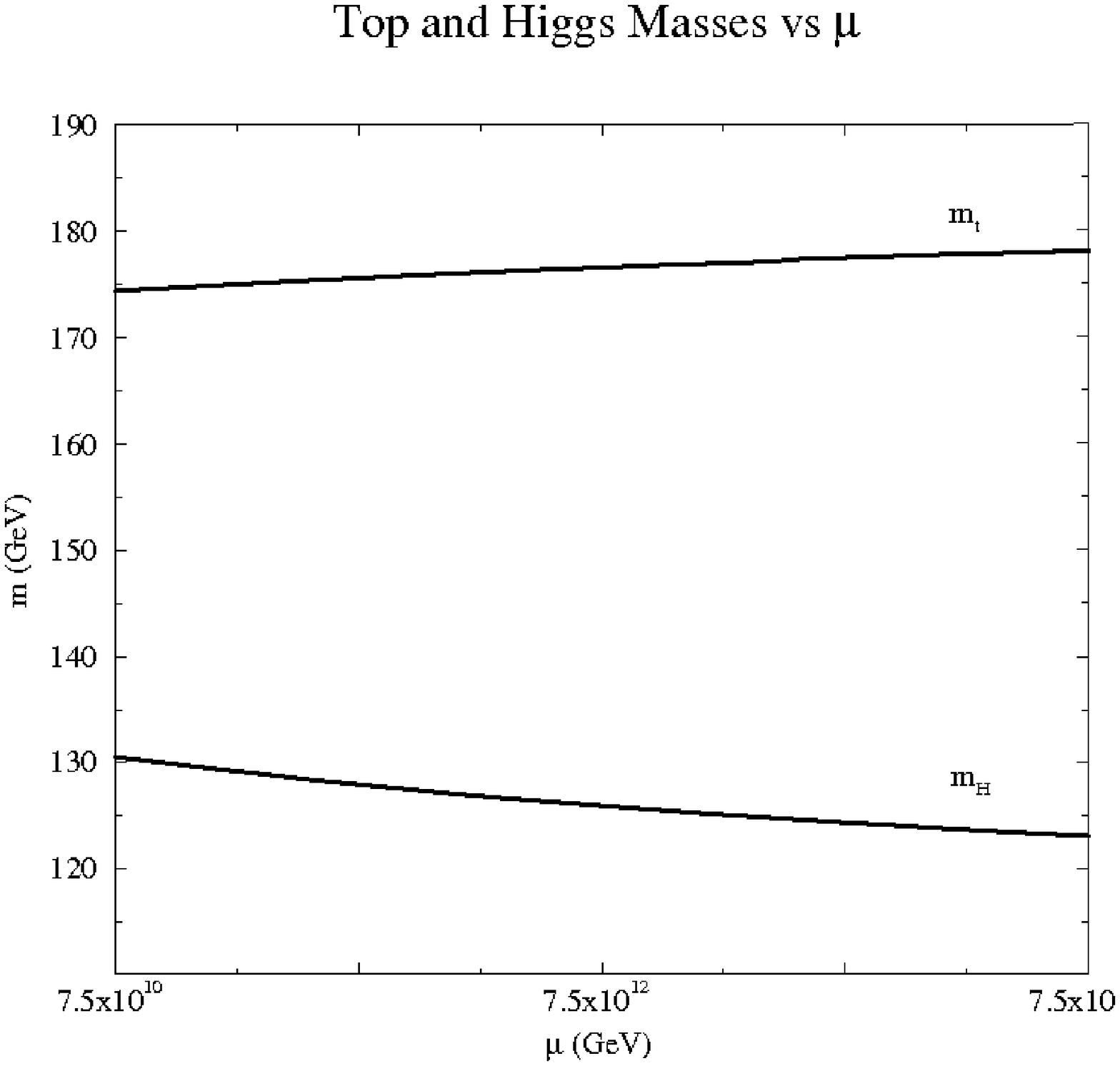}
\caption[]{Variation of the top and Higgs masses with the matching scale $\mu$ over
a range from $10^{10}$ to $10^{14}$ GeV. The scale $\mu=7.5\times 10^{12}$,
for which $\sin^2(\theta(\mu))=\frac{3}{8}$, is defined as a
`central value'.}
\label{fig:Fig.(2)}
\end{figure}
\clearpage

\begin{figure}
\includegraphics*[scale=0.8]{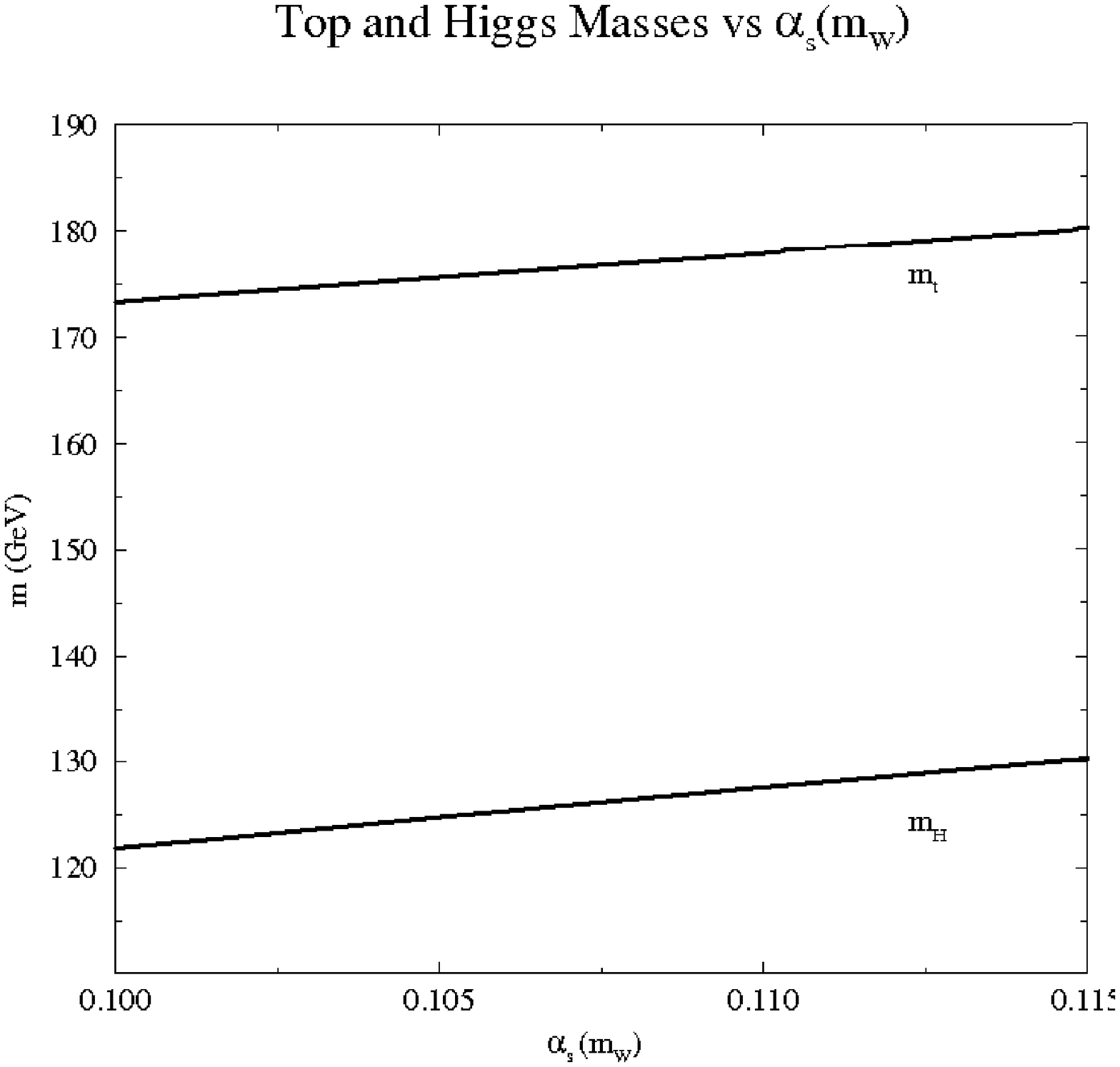}
\caption[]{Variation of $m_t$ and $m_H$ with the strong coupling; $\alpha_S=0.107$ is
considered the central value.}
\label{fig:Fig.(3)}
\end{figure}
\clearpage

\begin{figure}
\includegraphics*[scale=0.8]{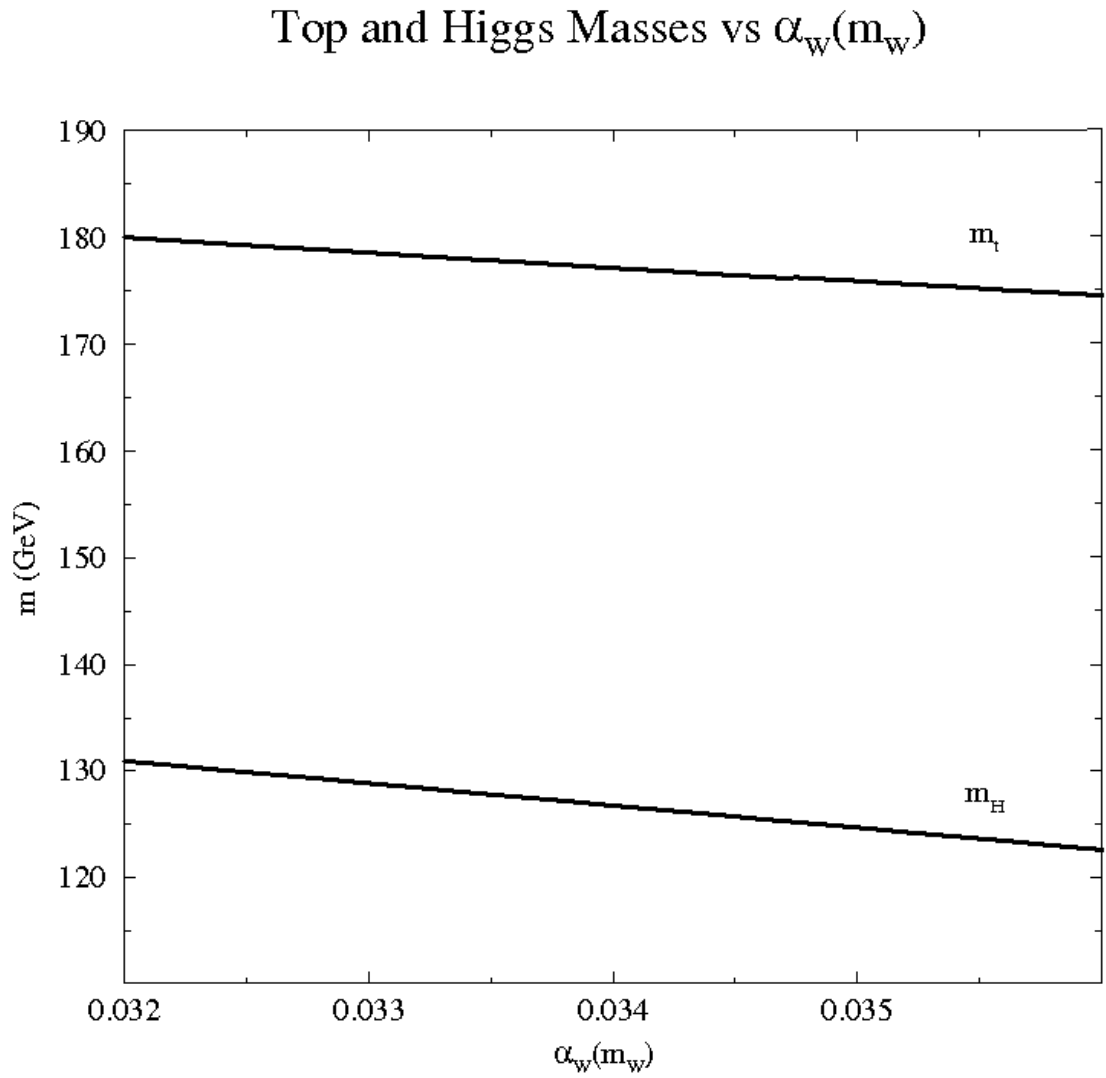}
\caption[]{Variation of $m_t$ and $m_H$ with the weak coupling;  $\alpha_W=0.0344$
is the central value.}
\label{fig:Fig.(4)}
\end{figure}
\clearpage

\begin{figure}
\includegraphics*[scale=1.0]{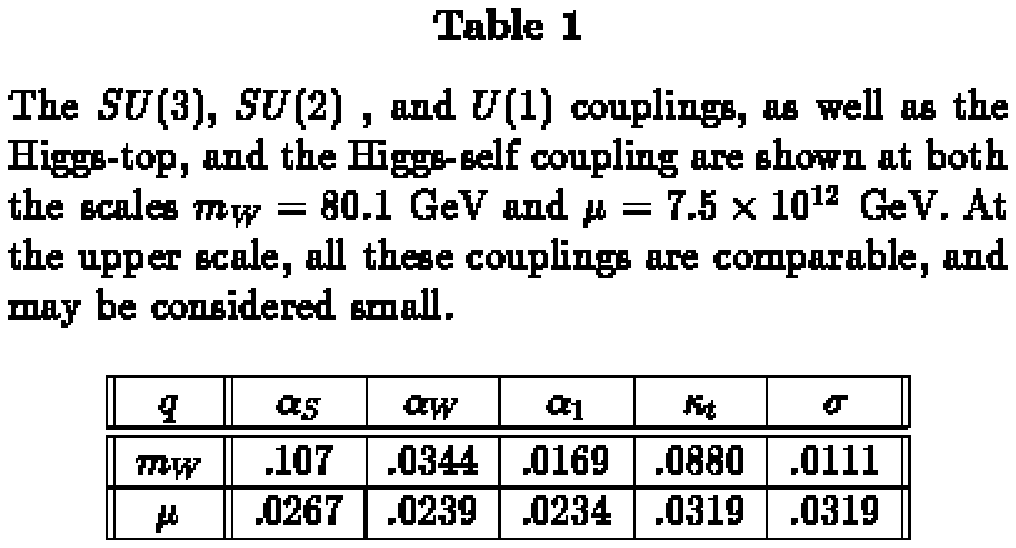}
\label{fig:Fig.(5)}
\end{figure}
\clearpage

\end{document}